\documentclass[10pt,twocolumn,letterpaper]{article}

\usepackage{cvpr}
\usepackage{times}
\usepackage{epsfig}
\usepackage{graphicx}
\usepackage{amsmath}
\usepackage{amssymb}
\usepackage{float}
\usepackage{stfloats}

\usepackage[breaklinks=true,bookmarks=false]{hyperref}

\title{Neural Differential Equations for Single Image Super-Resolution}

\author{
Teven Le Scao \\
Hugging Face \thanks{Work performed while at the Czech Technical University in Prague} \\
{\tt teven@huggingface.co}
}

\cvprfinalcopy 

\ifcvprfinal\fi

\title{Neural Differential Equations for Single Image Super-Resolution}

\begin{document}

\maketitle

\begin{abstract}
 Although Neural Differential Equations have shown promise on toy problems such as MNIST, they have yet to be successfully applied to more challenging tasks. Inspired by variational methods for image restoration relying on partial differential equations, we choose to benchmark several forms of Neural DEs and backpropagation methods on single image super-resolution. The adjoint method previously proposed for gradient estimation has no theoretical stability guarantees; we find a practical case where this makes it unusable, and show that discrete sensitivity analysis has better stability. In our experiments\footnote{Code available at {\tt github.com/TevenLeScao/BasicSR}}, differential models match the performance of a state-of-the art super-resolution model.
\end{abstract}

\section{Introduction}

A growing body of work is exploring the connection between deep learning and dynamical systems. Historically, the objective has been to help solve differential equations with deep learning \cite{GONZALEZGARCIA1998S965}. Inversely, recent work has borrowed from dynamical systems literature to train neural networks \cite{Nesterov}. Neural Ordinary Differential Equations \cite{NIPS2018_7892} are such an approach that approximates functions with a differential equation parameterized by a neural network. They take the input as the boundary condition of an initial value problem and integrate the equation to produce the output.

This paper attempts to benchmark different forms of Neural ODEs, which have mostly been tested on early image classification datasets such as MNIST or CIFAR for now. Since good performance there does not necessarily transfer to harder problems, we use supervised single image super-resolution (SR). We choose SR as partial differential equations (PDEs) have been used on that task \cite{PDESR}\cite{TNRD}. We can then train a Neural ODE parameterized by a convolutional neural network to learn the discretized form of an adequate PDE, considering the action of the CNN like a finite difference approximation of the equation.

\begin{figure}
    \centering
    \includegraphics[width=\columnwidth]{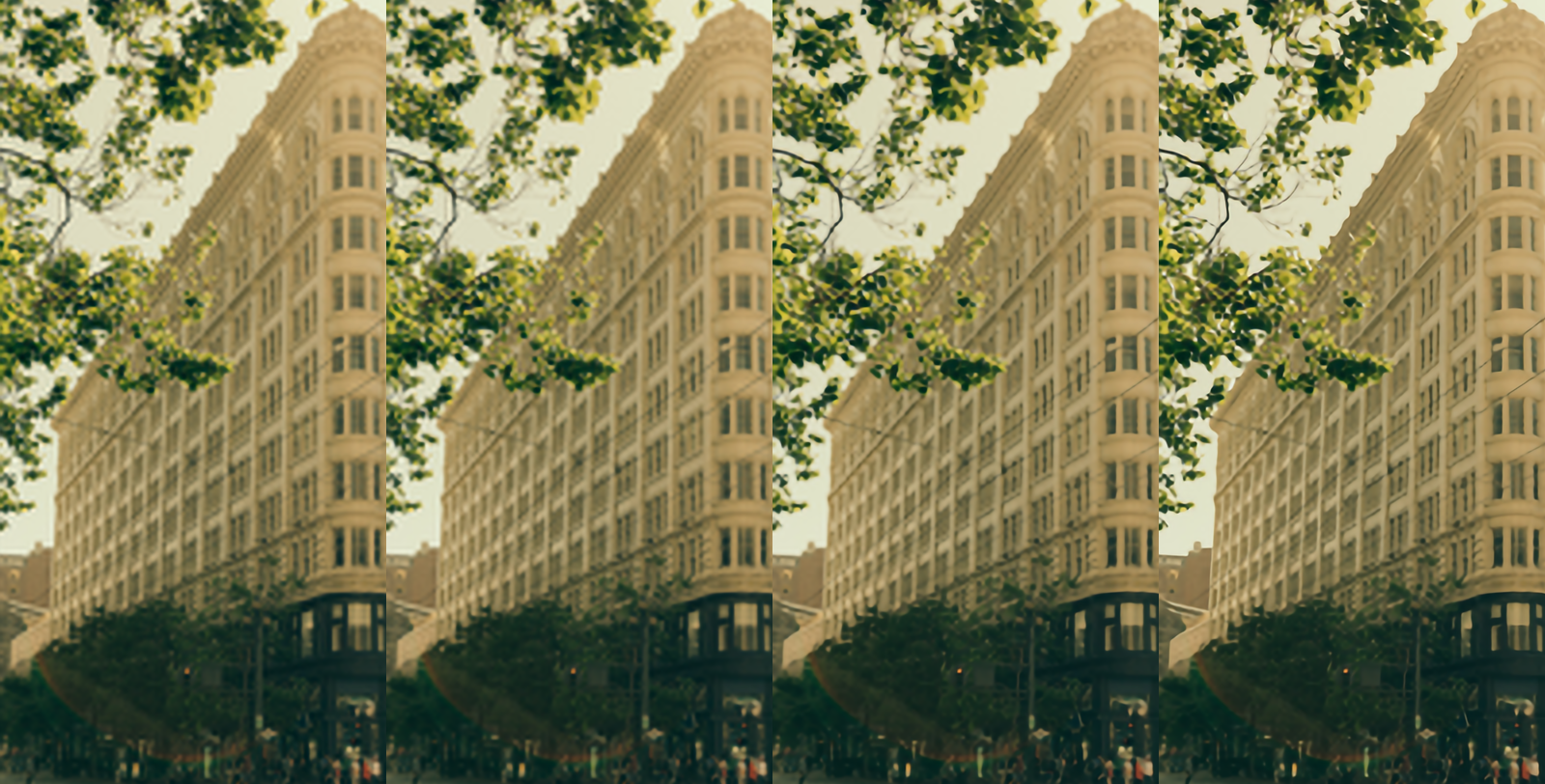}
    \caption{\label{fig : evolution} Our model learns a partial differential equation parameterized by a neural network that smoothly transforms the low-resolution input (left) into a high-resolution reconstruction (right).}
\end{figure}

Our contributions are threefold. First, we test Neural ODEs variants and corresponding optimization techniques \cite{Augmented}\cite{ANODE}. We find that dimension augmentation and time-dependence have positive impacts, and that discrete reverse sensitivity analysis is significantly more stable than the adjoint method advocated in \cite{NIPS2018_7892}. Then, we compare Neural ODEs with the residual networks usually employed for super-resolution. We find they're competitive with state-of-the art systems while using a fraction of the parameter size, although this does not translate to a speed increase. Finally, we investigate the benefits of their computational adaptivity, and observe a correlation between the amount of computation they use on an image and the performance gain of residual networks on that image as they grow deeper.

\section{Related work}

\textbf{Supervised single-image super-resolution} is a task where low- and high-resolution image pairs are provided as training data to learn a mapping from the low-resolution space to the high-resolution space. Residual networks \cite{Residual} are typically used. Those are trained with a combination of pixel-wise loss, image classifier features \cite{FeatureLoss1} (those should be the same for the generated image and the original) and adversarial networks \cite{SRGAN}. We replace residual generators with Neural ODEs, and stay in the non-adversarial case.

There has been extensive work on the relationship between \textbf{residual networks and dynamical systems}. Most of it leverages the connection to build principled residual architectures \cite{BeyondFinite}\cite{ResidualDynamical}\cite{MotivatedResnets}\cite{RungeKuttaCNN}. In contrast, Neural ODEs \cite{NIPS2018_7892} introduce a concurrent architecture that continuously deforms inputs according to a differential equation for a fixed amount of integration time to produce the output. We study its behaviour and performance on a real-world task.

Image restoration tasks, such as denoising \cite{PDEdenoising}, deblurring \cite{PDEdeblurring}, inpainting \cite{PDEinpainting} have historically been solvable with \textbf{total variation methods} \cite{TotalVariation} which naturally lead to PDEs. Super-resolution has also used differential methods \cite{TNRD}\cite{PDESR}, so Neural ODEs, which learn differential equations to fit the data in a supervised way, are a promising class of models for the task. Previous work that learns equation parameters \cite{TNRD} typically uses constraints on the filters in order to leverage the exact equivalence between convolution filters and spatial differentiation \cite{ConvolutionPDE}. In contrast, we use a neural network approach where we overspecify the model and use larger datasets and limited a priori knowledge.

\section{Approach}

\subsection{Architecture}

\begin{figure*}[b]
    \centering
    \includegraphics[width=0.85\textwidth]{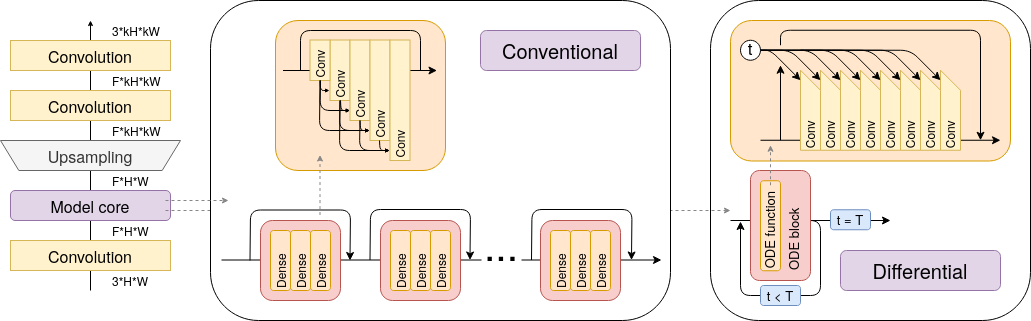}
    \caption{The super-resolution network. A first convolution extends the number of channels to $F$ filters. Then, a model core performs feature extraction in filter space. This feature map is upsampled to the desired scale, then a set of convolutions produces the final.}
    \label{fig : archi}
\end{figure*}

We use as reference the residual-in-residual-dense-blocks (RRDB) network \cite{ESRGAN}, a state-of-the art architecture that has proved effective for both adversarial and non-adversarial training. We replace the main residual module with a differential module as shown in fig. \ref{fig : archi}. Inputs go through a first convolution to expand the RGB channels to $F$ filters. Then, a residual network or Neural ODE iterates over the image in filter space to perform feature extraction. Finally, after upsampling to the final resolution and applying a convolution in filter space, the output convolution returns to RGB channel space to yield the image. We use LeakyRELU activations and no batch normalization.

In conventional models, the model core consists of hundreds of convolution layers arranged in an intricate architecture. In RRDB networks, the basic unit is a dense block of five convolutions with a residual connection bypassing it. Those are then arranged into groups of three units with residuals. The final model contains a pre-specified number of those groups: 5 in our low-data experiment, 20 in the high-data one, for a total of 75 to 300 layers of varying sizes. In contrast, the Neural ODE uses only one set of stacked convolutions, which we call the ODE function: 2 in the low-data regime and 7 in the high-data one. We initialize a time variable $t$ at $t_0 = 0$ and an ODE state $u_0$ at the input value. At every step $k$, the ODE solver takes in the current state $u_k \approx u(t_k)$ and computes the time increment $\Delta t$ and state increment $\Delta u$ according to the ODE function. The exact calculation depends on the solver; here, the Dormand-Prince method makes several function calls between $t$ and $ + \Delta t$. The integration then proceeds forward with $t_{k+1} = t + \Delta t$ and $u_{k+1} = u_k + \Delta u$, until we reach a pre-specified $t = T_{final}$, at which point the ODE block outputs the final value $u(T)$ for the rest of the network. We concatenate a time channel that contains the current integration time $t$ to the filter channels to model time dependence \cite{NIPS2018_7892}. The differential system uses considerably less parameters: our high-data differential image generator, for example, is 25 times lighter than its conventional equivalent.

Finally, we also experiment with dimension augmentation \cite{Augmented}, where extra dimensions are concatenated to the latent space in order to allow possible trajectories that would have to cross in the lower-dimensional space and ANODEs \cite{ANODE}, which implement a checkpointing scheme to ensure the stability of the gradient signal.

\subsection{Backpropagation methods}  \label{Optimization}

A Neural ODE attempts to map inputs $\mathbf{x}(0)$ to outputs $\mathbf{x}(T)$ that approximate target $\mathbf{y}$ by the continuous action of dynamical system $\dot{\mathbf{x}} = f(\mathbf{x}, t, \theta)$, with $f$ a neural network of parameters $\theta$. Training it is finding the minimum of
\begin{equation}
loss(\theta) = d(\mathbf{y}, \mathbf{x}(0) + \int_{0}^{T}f(\mathbf{x}(t), t, \theta)dt)
\end{equation}
with regard to $\theta$, where $d$ is the distance function we're using for training and $T$ is an arbitrary integration time. In order to use gradient descent, we must derive through the integral. Fortunately, if we consider this as an optimal control problem, we can use use sensitivity analysis to compute the derivative of the integral term in (1) with regard to $\theta$. 

When the number of parameters of the model is large compared to the dimension of the space of the differential equation, as here, adjoint sensitivity is the most efficient technique \cite{FATODE}. It first computes the trajectory of the system normally, then solves another differential equation system, the adjoint, that computes the derivatives backwards in time along that trajectory. It is the method used in \cite{NIPS2018_7892} which allows for constant memory cost as a function of depth when used without checkpointing, ie saving intermediary values of the forward pass. However, in this case, due to error accumulation, the trajectory of the adjoint equation may completely diverge from the trajectory of the forward pass if there's no additional stability guarantees (for example, a reversible integrator, which may have worse accuracy).

Another possibility is discrete reverse sensitivity: simply propagating the gradients through the internal operations of the solver using the chain rule. This is equivalent to backpropagation, or, in automatic differentiation terms, reverse accumulation. It requires access to the solver operations, rather than treating it like a black box like the adjoint method does, and does not seem to scale as well on performance benchmarks \cite{JuliaSensitivityAnalysis}. However, it does not have the instability risk of the adjoint method. As most of the work comparing those methods has been performed on problems stemming from physical systems, which might not be representative of the equations that appear in machine learning applications, we will test and compare both methods.

\section{Experiments}

We conducted experiments using five usual super-resolution datasets: BSD \cite{BSDS} and DIV2K \cite{DIV2K} for training, and Urban 100, General 100 and Set 14 for testing. Pre-processing measures and hyperparameters are presented in appendix \ref{appendixhyper}, and visual results in appendix \ref{appendiximages}.

\subsection{Architecture search on the BSD dataset}

As this smaller dataset allows an extensive grid architecture search, we identify the best-performing differential system on the BSD dataset before moving on to DIV2K. We train baseline neural ODEs (NODEs), augmented ODEs and ANODEs, all with an ODE function of two convolution layers. Wherever possible, we test time-dependent (e.g. concatenating a channel with the current integration time to the image tensor to model a time-dependent PDE) and  autonomous (e.g. no time dependence) ODE blocks. Finally, both discrete and adjoint training are presented. We compare with the state-of-the-art RRDBNet generator from \cite{ESRGAN}. Table \ref{tab : BSD results} presents test PSNRs for each system.

\begin{table}
\centering
\caption{ \label{tab : BSD results}PSNR results training on BSD and testing on the concatenation of all datasets. Time-dependency, adjoint optimization, and dimension augmentation have a generally positive effect. Augmented time-dependent ODE are the best-performing models, with the adjoint-optimized version beating RRDB by 0.43 PSNR. Results are presented split by dataset in appendix \ref{split}.}
\vspace{0.5cm}
\begin{tabular}{|l|c|c|}
\hline
\ \ \textbf{Architecture} & \textbf{Discrete} & \textbf{Adjoint}
\\ \hline
\ \ RRDBNet \cite{ESRGAN} & 25.33 & -
\\ \hline
\ \ ANODE \cite{ANODE} & 24.40 & -
\\ \hline
 \begin{tabular}{ll}NODE & \begin{tabular}{@{}c@{}}Autonomous \\ Time-dependent\end{tabular}\end{tabular} & \begin{tabular}{@{}c@{}}24.15 \\ 25.67\end{tabular} & \begin{tabular}{@{}c@{}} 25.57 \\ 25.61 \end{tabular}
\\ \hline
 \begin{tabular}{ll}Augmented \cite{Augmented} & \begin{tabular}{@{}c@{}}Aut. \\ Time-dep.\end{tabular}\end{tabular} & \begin{tabular}{@{}c@{}}25.24 \\ 25.19\end{tabular} & \begin{tabular}{@{}c@{}}25.67 \\ \textbf{25.76}\end{tabular}
\\ \hline
\end{tabular}
\end{table}

In this low-data regime, differential systems are competitive against RRDBNet, with Augmented time-dependent ODEs beating it by an encouraging 0.10 PSNR, although the visual quality difference is minimal. We attribute the slightly lower PSNR of the conventional model mostly to overfitting, as higher validation scores for the state-of-the-art CNN do not translate to better PSNR on the testing dataset, Urban 100. In contrast, those are consistent for the augmented time-dependent ODE. Finally, adjoint optimization does not seem to have a consistent effect. Baseline NODEs may lead to less stable adjoint gradients than Augmented ODEs, explaining the difference in performance.

\subsection{Full training on the DIV2K dataset}

\begin{table}[b]
\centering
\caption{ \label{tab : DIV results}PSNR results training on the DIV2K dataset. Training with the adjoint method is intractable.}
\vspace{0.5cm}
\begin{tabular}{|l|c|c|}
\hline
\textbf{Architecture} & \textbf{Discrete} & \textbf{Adjoint}
\\ \hline
RRDBNet & \textbf{26.88} & -
\\ \hline
Augmented time-dependent ODE & 26.69 & \textit{divergent}
\\ \hline
\end{tabular}
\end{table}

We keep using the augmented time-dependent architecture. As in the low-data regime, the difference in PSNR is not noticeable without zooming in. Images with a significant difference in score one way or another tend to be hard images with black-and-white striped patterns, where the worse-performing model generates stripes of the wrong color or not aligned, which causes a heavy PSNR penalty without influencing perceived image quality noticeably.

Notably, \textbf{no adjoint-optimized model trained successfully}. As mentioned in section \ref{Optimization}, for an arbitrary solving method and differential equation, there is no guarantee that the adjoint problem we need to solve in the backward pass is tractable. Previous work has trained image classifiers on MNIST or CIFAR without stability issues. However, in our experiments on this task with larger models, there is invariably a training batch for which it becomes intractable. Stability guarantees seem necessary to impose bounds on the number of steps the adjoint pass requires: we've observed the backward pass require up to $10^5$ function evaluations (i.e. model calls) before stopping for lack of time as this had made it around $2000$ times longer than usual.

\subsection{Adaptive computation} \label{NFE}

\begin{figure}
    \centering
    \includegraphics[width=0.8\columnwidth]{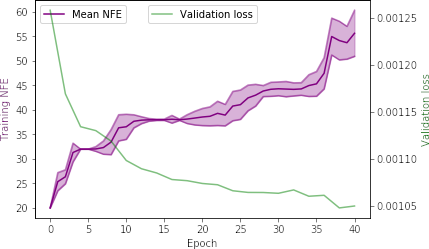}
    \caption{\label{NFEs_train} Evolution of the mean number of function evaluations per validation pass, with a band of a standard deviation.}
\end{figure}

One of the strategies of differential equation solvers that Neural ODEs aim to benefit from is adaptive computation. Indeed, modern solvers are able to vary the number of function evaluations (NFE) depending on the difficulty of the problem and the desired error tolerance. Previous work has found that NFE increases consistently as the model is training \cite{NIPS2018_7892} and interpreted it as proof that the modeled ODE gets harder as it learns to fit the data, although dimension augmentation reduces that effect \cite{Augmented}. Fig. \ref{NFEs_train} plots this increase over the training of our DIV2K differential model. At the end of training, NFE is up to between 50 and 80 evaluations (corresponding to 8 to 13 steps). The variance is especially interesting as we're using batches of 16 random patches, which we expected to have a homogeneizing effect.

In order to investigate whether the difficulty of the ODE correlates with the difficulty of the task, we look at the NFE for each model call rather than over whole epochs. In order to measure the difficulty of an image, we train four traditional residual models of varying (1, 2, 5, 10) depths in addition to our best performing 20-blocks RRDB network. One of the main premises of Neural ODEs is that additional solver method calls are equivalent to the model emulating a deeper conventional network. If that adaptive computation plays a role in ODE performance, there should be a link between which images prompt the solver to require more function evaluations and which images see the biggest performance increase with more CNN layers.

\begin{figure} []
    \centering
    \includegraphics[width=0.7\columnwidth]{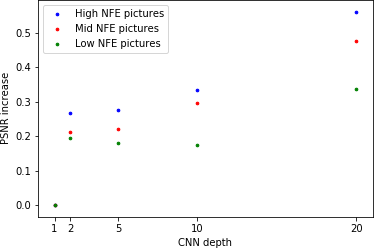}
    \caption{\label{NFEs_cascade} Performance increase with more blocks for the RRDB architecture on testing images that required low, medium, and high numbers of function evaluations. The increase in performance is higher on the pictures that required more function evaluations.}
\end{figure}

At test time, since a single image corresponds to around 40 patches of heterogeneous difficulty (as foreground tends to be easy and background to be hard), function evaluations are more clustered. We find that most (75 out of 100) images demand 8 solver steps, whereas 12 images require 7 steps and 13 images require 9. We dub those the medium, low and high NFE groups respectively and plot for each the increase in performance for RRDB as CNN depth increases in fig. \ref{NFEs_cascade}. As PSNR does not always increase with depth for the low NFE images while the high NFE pictures see it consistently increase, the differential equation solver does seem to require more calls in cases where it helps performance, although the spread is smaller than at training time. 

As shown in table \ref{tab : cascade table}, the differential model achieves similar performance to traditional residual networks with only a fraction of the parameters; however, its speed is comparable, especially at the end of training when it requires more function evaluations. In a sense, Neural ODEs can be seen as a hypernetwork \cite{Hyper}, with a limited set of parameters defining an extensive set of computations.

\begin{table}
\centering
\caption{ \label{tab : cascade table}Test PSNR performance, model size, and epoch time for various conventional networks and the ODE model.}
\vspace{0.5cm}
\begin{tabular}{|l|l|l|l|}
\hline
\textbf{Model} & \textbf{Parameters} & \textbf{Epoch time} & \textbf{PSNR}
\\ \hline
1 RRDB blocks & $0.87*10^6$ & 2mn 46s & 23.80
\\ \hline
2 RRDB blocks & $1.6*10^6$ & 3mn 20s & 24.02
\\ \hline
5 RRDB blocks & $3.7*10^6$ & 5mn 2s & 24.03
\\ \hline
10 RRDB blocks & $7.3*10^6$ & 8mn 4s & 24.09
\\ \hline
20 RRDB blocks & $15*10^6$ & 13mn 57s & \textbf{24.27}
\\ \hline
7-layer ODE & $0.57*10^6$ & 10-17mn & \textbf{24.10}
\\ \hline
\end{tabular}
\end{table}

\section{Conclusion}

We have presented a novel model for single image super-resolution that applies a neural differential equation instead of discrete pre-defined residual convolutional layers. We have also shown that the lack of stability guarantees of the continuous adjoint method materializes into practical intractability, and found that differential models use adaptive computation, allocating more function evaluations to problems that can benefit. This allowed us to match the performance a state-of-the art super-resolution with only a fraction of the parameters, although with similar time cost. 

\medskip

The author would like to thank Tom\'a\v{s} Mikolov, Josef \v{S}ivic, and Germ\'an Kruszewski for their guidance and advice. This work was supported by the European Regional Development Fund under the project IMPACT no. CZ.02.1.01/0.0/0.0/15 003/0000468.

\bibliography{egbib}

\begin{thebibliography}{10}\itemsep=-1pt

\bibitem{DIV2K}
Eirikur Agustsson and Radu Timofte.
\newblock {NTIRE} 2017 challenge on single image super-resolution: Dataset and
  study.
\newblock In {\em The IEEE Conference on Computer Vision and Pattern
  Recognition (CVPR) Workshops}, July 2017.

\bibitem{ConvolutionPDE}
Jian-Feng Cai, Bin Dong, Stanley Osher, and Zuowei Shen.
\newblock Image restoration: Total variation, wavelet frames, and beyond.
\newblock {\em Journal of the American Mathematical Society}, 25(4):1033--1089,
  2012.

\bibitem{ResidualDynamical}
Bo Chang, Lili Meng, Eldad Haber, Frederick Tung, and David Begert.
\newblock Multi-level residual networks from dynamical systems view.
\newblock In {\em International Conference on Learning Representations}, 2018.

\bibitem{NIPS2018_7892}
Tian~Qi Chen, Yulia Rubanova, Jesse Bettencourt, and David~K Duvenaud.
\newblock Neural ordinary differential equations.
\newblock In S. Bengio, H. Wallach, H. Larochelle, K. Grauman, N. Cesa-Bianchi,
  and R. Garnett, editors, {\em Advances in Neural Information Processing
  Systems 31}, pages 6571--6583. Curran Associates, Inc., 2018.

\bibitem{PDESR}
Yuanxu Chen, Yupin Luo, and Dongcheng Hu.
\newblock {A general approach to blind image super-resolution using a {PDE}
  framework}.
\newblock In Shipeng Li, Fernando Pereira, Heung-Yeung Shum, and Andrew~G.
  Tescher, editors, {\em Visual Communications and Image Processing 2005},
  volume 5960, pages 1819 -- 1830. International Society for Optics and
  Photonics, SPIE, 2005.

\bibitem{TNRD}
Y. {Chen} and T. {Pock}.
\newblock Trainable nonlinear reaction diffusion: A flexible framework for fast
  and effective image restoration.
\newblock {\em IEEE Transactions on Pattern Analysis and Machine Intelligence},
  39(6):1256--1272, June 2017.

\bibitem{FeatureLoss1}
Alexey Dosovitskiy and Thomas Brox.
\newblock Generating images with perceptual similarity metrics based on deep
  networks.
\newblock {\em CoRR}, abs/1602.02644, 2016.

\bibitem{Augmented}
Emilien Dupont, Arnaud Doucet, and Yee~Whye Teh.
\newblock Augmented neural odes.
\newblock {\em arXiv preprint arXiv:1904.01681}, 2019.

\bibitem{ANODE}
Amir Gholaminejad, Kurt Keutzer, and George Biros.
\newblock {ANODE}: Unconditionally accurate memory-efficient gradients for
  neural odes.
\newblock In {\em IJCAI}, 2019.

\bibitem{GONZALEZGARCIA1998S965}
R. Gonz\'alez-Garc\'ia, R. Rico-Mart\'inez, and I.G. Kevrekidis.
\newblock Identification of distributed parameter systems: A neural net based
  approach.
\newblock {\em Computers and Chemical Engineering}, 22:S965 -- S968, 1998.
\newblock European Symposium on Computer Aided Process Engineering-8.

\bibitem{Hyper}
David Ha, Andrew~M. Dai, and Quoc~V. Le.
\newblock Hypernetworks.
\newblock {\em CoRR}, abs/1609.09106, 2016.

\bibitem{Residual}
Kaiming He, Xiangyu Zhang, Shaoqing Ren, and Jian Sun.
\newblock Deep residual learning for image recognition.
\newblock {\em CoRR}, abs/1512.03385, 2015.

\bibitem{SRGAN}
Christian Ledig, Lucas Theis, Ferenc Huszar, Jose Caballero, Andrew~P. Aitken,
  Alykhan Tejani, Johannes Totz, Zehan Wang, and Wenzhe Shi.
\newblock Photo-realistic single image super-resolution using a generative
  adversarial network.
\newblock {\em CoRR}, abs/1609.04802, 2016.

\bibitem{BeyondFinite}
Yiping Lu, Aoxiao Zhong, Quanzheng Li, and Bin Dong.
\newblock Beyond finite layer neural networks: Bridging deep architectures and
  numerical differential equations.
\newblock {\em CoRR}, abs/1710.10121, 2017.

\bibitem{BSDS}
D. Martin, C. Fowlkes, D. Tal, and J. Malik.
\newblock A database of human segmented natural images and its application to
  evaluating segmentation algorithms and measuring ecological statistics.
\newblock In {\em Proc. 8th Int'l Conf. Computer Vision}, volume~2, pages
  416--423, July 2001.

\bibitem{Nesterov}
Michael Muehlebach and Michael~I. Jordan.
\newblock A dynamical systems perspective on {Nesterov} acceleration.
\newblock In {\em Proceedings of the 36th International Conference on Machine
  Learning, {ICML} 2019, 9-15 June 2019, Long Beach, California, {USA}}, pages
  4656--4662, 2019.

\bibitem{JuliaSensitivityAnalysis}
Christopher Rackauckas, Yingbo Ma, Vaibhav Dixit, Xingjian Guo, Mike Innes,
  Jarrett Revels, Joakim Nyberg, and Vijay Ivaturi.
\newblock A comparison of automatic differentiation and continuous sensitivity
  analysis for derivatives of differential equation solutions.
\newblock {\em CoRR}, abs/1812.01892, 2018.

\bibitem{PDEdeblurring}
Leonid Rudin, Pierre-Luis Lions, and Stanley Osher.
\newblock {\em Multiplicative Denoising and Deblurring: Theory and Algorithms},
  pages 103--119.
\newblock Springer New York, New York, NY, 2003.

\bibitem{TotalVariation}
Leonid~I. Rudin, Stanley Osher, and Emad Fatemi.
\newblock Nonlinear total variation-based noise removal algorithms.
\newblock {\em Physica D: Nonlinear Phenomena}, 60(1):259 -- 268, 1992.

\bibitem{MotivatedResnets}
Lars Ruthotto and Eldad Haber.
\newblock Deep neural networks motivated by partial differential equations.
\newblock {\em CoRR}, abs/1804.04272, 2018.

\bibitem{PDEinpainting}
Carola-Bibiane Schönlieb.
\newblock {\em Partial Differential Equation Methods for Image Inpainting}.
\newblock Cambridge Monographs on Applied and Computational Mathematics.
  Cambridge University Press, 2015.

\bibitem{PDEdenoising}
{Seongjai Kim}.
\newblock {PDE}-based image restoration: a hybrid model and color image
  denoising.
\newblock {\em IEEE Transactions on Image Processing}, 15(5):1163--1170, May
  2006.

\bibitem{ESRGAN}
Xintao Wang, Ke Yu, Shixiang Wu, Jinjin Gu, Yihao Liu, Chao Dong, Chen~Change
  Loy, Yu Qiao, and Xiaoou Tang.
\newblock {ESRGAN:} enhanced super-resolution generative adversarial networks.
\newblock {\em CoRR}, abs/1809.00219, 2018.

\bibitem{FATODE}
Hong. Zhang and Adrian. Sandu.
\newblock {FATODE}: A library for forward, adjoint, and tangent linear
  integration of {ODEs}.
\newblock {\em SIAM Journal on Scientific Computing}, 36(5):C504--C523, 2014.

\bibitem{RungeKuttaCNN}
Mai Zhu and Chong Fu.
\newblock Convolutional neural networks combined with {Runge-Kutta} methods.
\newblock {\em CoRR}, abs/1802.08831, 2018.

\end{thebibliography}
\bibliographystyle{ieee_fullname}

\appendix

\section{Datasets, evaluation and training procedures} \label{appendixhyper}

The BSD dataset was originally gathered for image segmentation tasks and contains 300 photos of variable, small size, around 400x400 pixels large. Those are separated into 200 training images and 100 testing ones. After using bicubic downsampling to generate factor 4x low-resolution images and dividing the images into smaller patches for training following the pre-processing pipeline in \cite{ESRGAN}, we obtain a training set of 1200 patches. By contrast, the DIV2K dataset is much richer; it contains 900 images of higher starting resolution, separated in 800 training and 100 validation images. We apply the same process as previously and obtain a training set of 33152 patches, 28 times bigger. The testing datasets are not pre-processed.

In order to present fair comparisons with the state of the art, we re-use many of the hyperparameters of \cite{ESRGAN} for training, such as learning rates ($2.10^{-4}$), number of filters (64), weight decay (none), and the image pre-processing procedure parameters. We differ by implementing validation-based learning rate decay: if a number (3 in our case) of validation results do not improve over the previous best result, we reduce the learning rate, and stop training when it reaches a minimum value. We use the default \textit{torchdiffeq} solving parameters.

\section{Low-data results split by dataset} \label{split}

\begin{table*}[b]
\centering
\caption{ \label{tab : BSD split results}PSNR results training on BSD split by dataset. Performance is consistent across datasets.}
\vspace{0.5cm}
\begin{tabular}{|l|ccc|c|}
\hline
\ \ \textbf{Architecture} & \textbf{Set 14} & \textbf{General 100} & \textbf{Urban 100} & \textbf{Concatenation}
\\ \hline
\ \ RRDBNet \cite{ESRGAN} & 25.62 & 27.84 & 22.79 & 25.33
\\ \hline
\ \ ANODE \cite{ANODE} & 24.77 & 26.78 & 21.97 & 24.40
\\ \hline
 \begin{tabular}{ll}NODE & \begin{tabular}{@{}c@{}}\begin{tabular}{ll}Autonomous & \begin{tabular}{@{}c@{}}Discrete \\ Adjoint\end{tabular}\end{tabular} \\ \begin{tabular}{ll}Time-dependent & \begin{tabular}{@{}c@{}}Discrete \\ Adjoint\end{tabular}\end{tabular}\end{tabular}\end{tabular} & \begin{tabular}{@{}c@{}c@{}c@{}} 24.55 \\ 25.82 \\ 25.89 \\ 25.85 \end{tabular} & \begin{tabular}{@{}c@{}c@{}c@{}} 26.45 \\ 28.11 \\ 28.21 \\ 28.15 \end{tabular} & \begin{tabular}{@{}c@{}c@{}c@{}} 21.79 \\ 23.00 \\ 23.10 \\ 23.03 \end{tabular} & \begin{tabular}{@{}c@{}c@{}c@{}} 24.15 \\ 25.57 \\ 25.67 \\ 25.61 \end{tabular} 
\\ \hline
 \begin{tabular}{ll}Augmented & \begin{tabular}{@{}c@{}}\begin{tabular}{ll}Autonomous & \begin{tabular}{@{}c@{}}Discrete \\ Adjoint\end{tabular}\end{tabular} \\ \begin{tabular}{ll}Time-dependent & \begin{tabular}{@{}c@{}}Discrete \\ Adjoint\end{tabular}\end{tabular}\end{tabular}\end{tabular} & \begin{tabular}{@{}c@{}c@{}c@{}} 25.47 \\ 25.88 \\ 25.47 \\ \textbf{25.96} \end{tabular} & \begin{tabular}{@{}c@{}c@{}c@{}} 27.76 \\ 28.20 \\ 27.60 \\ \textbf{28.30} \end{tabular} & \begin{tabular}{@{}c@{}c@{}c@{}} 22.69 \\ 23.11 \\ 22.75 \\ \textbf{23.20} \end{tabular} & \begin{tabular}{@{}c@{}c@{}c@{}} 25.24 \\ 25.67 \\ 25.19 \\ \textbf{25.76} \end{tabular} 
\\ \hline
\end{tabular}
\end{table*}

\section{Image results and comparisons} \label{appendiximages}

\begin{figure*}
    \centering
    \includegraphics[width=0.85\textwidth]{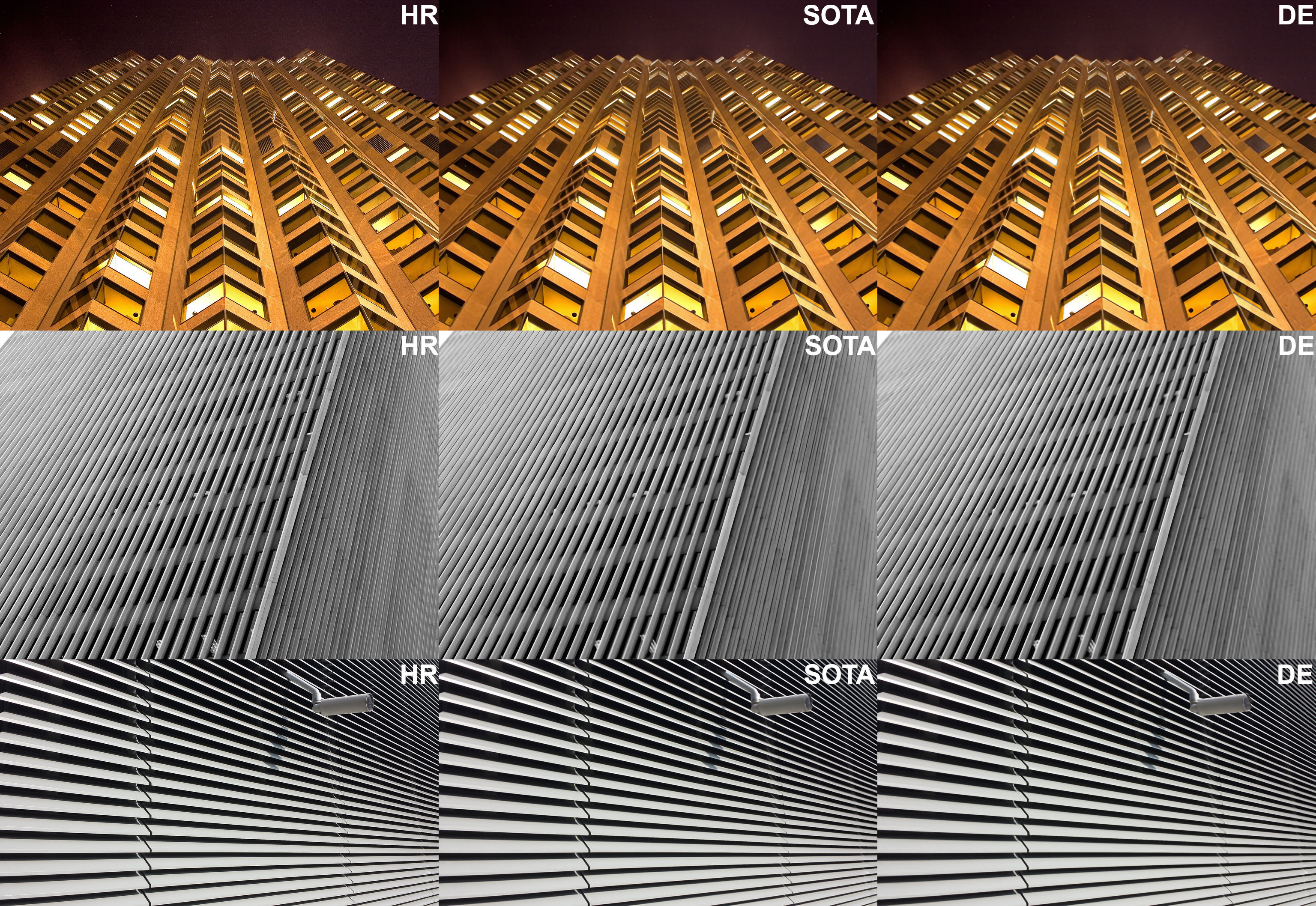}
    \caption{\label{comparison_DIV_fullimg} A comparison between images produced by the best-performing differential (center) and state-of-the-art (right) models and their high-resolution target (left). The first row shows an image where both models have typical PSNRs (24.28 for the ODE, 24.47 for the state-of-the-art). The second row shows the image where the ODE outperforms the state-of-the-art the most (1.47 PSNR difference). The third row shows the image where the state-of-the-art outperforms the ODE the most (1.08 difference) High score discrepancy images on both sides are hard black-and-white striped images where minute anti-aliasing or alignment differences are heavily penalized; this high variation in PSNR does not correlate well with human perception.}
\end{figure*}



\end{document}